\documentclass[12pt]{article}     % Paper without sections
\begin{document}

\begin{titlepage}
\samepage{
\begin{center}
 {\Large \bf The pion nucleon coupling constant and the Goldberger-Treiman Relation\\}
\bigskip
 {\large D.V. Bugg$^1$, M.D. Scadron$^2$ \\}
\vspace{0.10in}
 {\it $^1$ Queen Mary, University of London, London E1\,4NS, UK\\
      $^2$ Physics Department, University of Arizona, Tucson, Arizona 85721, USA }
\end{center}
\bigskip
\bigskip
\begin{abstract}
{\rm The latest $\pi N$ elastic scattering data are re-analysed to determine
the coupling constant $g_c$ of the charged pion, using the dispersion
relation for the invariant amplitude $B^{(+)}$. Depending on the choice
of data-base, values $g^2_c/4\pi = 13.80$ to 13.65 are obtained with
errors of $\pm 0.12$.
We re-examine the well known discrepancy with the
Goldberger-Treiman relation. After allowing for the mass
dependence of the pion decay constant $f_\pi$, a $(2{-}3)\%$ discrepancy is
predicted, hence $g^2_c/4\pi = 13.74 \pm 0.10$ in the prior case.

The mass difference between charge states of $\Delta (1232)$ is
$M^0 -M^{++} = 2.0 \pm 0.4$ MeV, close to twice the mass difference
between neutron and proton.
The difference in widths on resonance is $\Gamma ^0 - \Gamma ^{++}
=3.8 \pm 1.0$ MeV.
One may account for a width difference of 4.5 MeV
from phase space for decays and the extra channel $\Delta ^0 \to
\gamma n$. }
\end{abstract}
\vfill
\smallskip}
\end{titlepage}

\section {Introduction}
There have been long-running arguments over the magnitude of the
pion-nucleon coupling constant.
It appears that this argument can at last be resolved.
The history will be reviewed briefly below, so as to set the
present work in context.
The dispute arises mainly from discrepancies in the
normalisation of $d\sigma ^\pm /d\Omega$ data and total cross
sections in the region below the peak of the $\Delta (1232)$.
More recent data clarify the experimental situation.
The objective of the present paper is to make a fresh
determination of the pion nucleon coupling constant
$g^2_{\pi N\bar N}/4\pi$ with careful attention to
(i) Coulomb barrier corrections,
(ii) mass and width differences between $\Delta ^{++}$ and $\Delta ^0$.
Hopefully, this re-analysis will settle at least some of the
disagreements which have persisted for many years.

The Goldberger-Treiman relation [1] states that $f_\pi g_{\pi N\bar N}
\simeq m_ng_A$, where $m_n$ is the mean mass of the nucleon, 938.9 MeV,
$g_A$ the axial coupling constant of nucleon $\beta$ decay,
$1.267 \pm 0.004$, and $f_\pi$ the pion decay constant
$f_\pi  = 92.42 \pm 0.26$ MeV [2].
There is a well known discrepancy $\Delta = 2-3\%$ with this relation:
\begin {equation}
\Delta = 1 - \frac {m_ng_A}{f_\pi g_c},
\end {equation}
when one uses $g_c$, the coupling constant of charged pions to the
proton; we shall distinguish later the coupling constant $g_0$ for
neutral pions.
We shall re-examine this discrepancy.
The value of $g_c$ is determined at the pole, $q^2 = 0$,
whereas the experimental value of $f_\pi$ is determined at $q^2 =
m^2_\pi$. Today, the $q^2$ dependence of $f_\pi$ is accurately under
control from our understanding of Chiral Symmetry and how it is broken.

\section {A brief historical review}
Experiments at the CERN synchro-cyclotron in 1968-70 made precise
measurements of $\pi N$ scattering up to 290 MeV. Total cross
sections for $\pi ^\pm p$ were measured from 70 to 290 MeV, Carter et
al. [3]; Bussey et al. reported $d\sigma ^\pm/d\Omega$ from 88 to 292
MeV [4]; and the integrated cross section $\sigma ^0$ for charge
exchange from 90 to 290 MeV was measured by Bugg et al. [5]. A
partial wave analysis of these data was made [6] including the effects
of the Coulomb barrier and allowing for mass and width differences of
the $\Delta$; these differences in mass and width were conspicuous in
the total cross section data. A value $g^2_c/4\pi = 14.28 \pm  0.16$
was found using the unsubtracted dispersion relation for the $B^{+}$
amplitude [7].
This value depended significantly on additional data then available at
310 MeV. These have been superceded by a series of experiments at
PSI from 1975 to 1983 measuring differential cross sections
and polarisations for elastic scattering and charge exchange.
The effects of the new data were to reduce the width of the $\Delta$
slightly and hence reduce $g^2_c/4\pi$, though there was no
fresh analysis at the time.

The 1973 analysis used Coulomb barrier corrections determined by solving
a relativised Schro\" odinger equation [8]. A treatment of Coulomb
effects based on dispersion relations was made by Tromborg et al. in
1977 [9].

H\" ohler and collaborators carried out extensive analyses
of $\pi N$ elastic scattering up to $\sim 2$ GeV using dispersion
relations [10].
Comparisons are frequently made with the work of
Koch and Pietarinen [11].
However, it should be realised that this analysis omitted the
mass and width differences between $\Delta ^{++}$ and
$\Delta ^0$. Some of the discrepancies subsequently reported
between experiment and this analysis arise from this point.

Measurements were made of $d\sigma ^\pm /d\Omega$ at TRIUMF
by Brack et al. [12]. They reported substantially lower
normalisation than the results of Bussey et al. in the
mass range below 140 MeV. This discrepancy persists to this day.

De Swart and collaborators carried out a full analysis [13-16]
of $NN$ elastic scattering up to 350 MeV and reported
considerably lower values of $g^2/4\pi$.
Their 1993 values are $g^2_0/4\pi = 13.56 \pm 0.09$ for coupling to
$\pi ^0$ and $g^2_c/4\pi = 13.52 \pm 0.05$ for coupling to charged
pions [16].

Arndt and collaborators re-analysed $\pi N$ data using dispersion
relation constraints and found $g^2_c/4\pi = 13.75 \pm 0.15$ [17].
However, this analysis omitted the total cross section data of
Carter et al., and floated the normalisations of the Bussey et al.
differential cross sections. It also treated the Coulomb barrier
corrections in an approximate form. The main objective of the present
analysis is to restore the missing data and the full treatment
of the Coulomb barrier and see how much difference these make.

Recently, the Uppsala group has reported much higher values
of $g^2_c/4\pi$ from measurements of $np$ charge exchange differential
cross sections: $g^2_c/4\pi = 14.52 \pm 0.26$ [18].

Meanwhile there have been extensive measurements at PSI and
TRIUMF of differential cross sections and polarisations
for $\pi ^\pm p$ elastic scattering and charge exchange.
All published values up to 2002 are included here using the
SAID data base.
The PSI measurement of the $\pi ^-p$ scattering length
via pionic X-rays [19] is particulary important in providing
an anchor point for the $S$-wave amplitude at threshold.
It removes many of the uncertainties concerned with
$S$-waves at low energies.

\section {The $B^{(+)}$ dispersion relation}
For $\pi N$ scattering, the nucleon pole lies at $s = m^2_n$,
almost midway
between the physical regions for $\pi ^+p$ ($s >(m_p+m_\pi)^2)$ and $\pi
^-p$ ($s < (m_p-m_\pi ^2))$; here $m_\pi$ is the mass of the charged
pion.
The determination of the $\pi N\bar N$
coupling constant is then a matter of interpolation between
the physical regions for these two processes.
This interpolation is more stable than the extrapolation which
is required in $NN$ analyses from the physical region to the
pole at $t  = m^2_\pi$.
De Swart reports that, for the $NN$ case, the
coupling constant is determined mostly by low energy data;
the determination must therefore come mostly from the $u$-channel pole
below threshold at $s = 4m_n^2 - m^2_\pi$.

The $B^{(+)}$ dispersion relation contains a nucleon pole
term and an integral over the imaginary part of the amplitude.
In this particular combination of amplitudes, S-waves
are suppressed by a large factor and the imaginary part of
$B^{(+)}$ is dominated by $P_{33}$.
The total cross sections and normalisations of differential
cross sections therefore
play an essential role in determining $g^2_c$.

\section {Ingredients in the Analysis}
Our new analysis has been done using the SAID program [20] which
constrains the data in the energy range up to 800 MeV using fixed $t$
dispersion relations for $|t| = 0$ to 0.4 GeV$^2$.
However, we do not impose the GMO sum rule.
This relation relates the scattering length combination
$(a _1 - a_3)$ to the nucleon coupling constant and an integral
over total cross sections at all energies:
$$ J = \frac {1}{4\pi ^2} \int ^\infty _{(m_p+m_\pi)^2} dk (\sigma
^-_{tot} - \sigma ^+_{tot})/\nu ,$$
where $k$ is the lab momentum of the pion and $\nu$ its total lab
energy.
Bugg and Carter pointed out [21] that this integral is subject to
a sizeable correction for the effect of the Coulomb barrier, which
systematically enhances $\pi ^-p$ cross sections and suppresses
$\pi ^+p$.
Furthermore, there is the danger that errors in cross sections at
high energies bias the analysis of the region near threshold.

The formalism for Coulomb barrier corrections is described in
the 1973 analysis  of Carter et al. [6].
Here we try using as alternatives the
corrections evaluated by both Tromborg et al. [9] and
Bugg [8], in order to check the magnitude of any differences
between them.
There is a point here which deserves clarification.
A superficial reading of these two papers suggests very
different numerical values for  corrections to $\pi ^-p$.
However, the two analyses adopt somewhat different approaches.
The analysis of Tromborg et al. includes allowance in the
numerical values of Coulomb barrier factors $C_{ij}$ for the slightly
different final-state momenta between $\pi ^-p \to \pi ^-p$ and $\pi ^-
p \to \pi ^0n$. The partial wave analysis of Carter al al. instead
allows specific phase space differences in these channels and
accordingly introduces a small inelasticity into the $P_{33}$
amplitude for $\pi ^-p$ scattering. When the analysis is
run with the two alternative formalisms, results agree within
one standard deviation for $P_{33}$ and better than this
for other partial waves.
Numerical values of the Coulomb barrier calculations are available
up to 500 MeV. At higher energies, extended-source Coulomb
barrier factors supplied by Gibbs have been used [22].

There is a Coulomb term $C_{13}$ which allows for explicit mixing
between $I = 1/2$ and $3/2$; it arises from the fact that the
Coulomb potential acts in the $\pi ^-p$ channel but not in
$\pi ^0n$. Including $C_{13}$ into the analysis improves
$\chi ^2$ significantly, by $\sim 100$.
More exactly, the fit has 21786 degrees of freedom; without $C_{13}$,
$\chi ^2 = 44850$ and including it $\chi ^2 \to 44752$. It
has the effect of increasing $g^2_c/4\pi$ by 0.03.

Our analysis also includes a mass difference between $\Delta
^{++}$ and $\Delta ^0$. This turns out to be essential.

There is an important detail concerning the total cross sections
of Carter et al. There was an uncertainty of $\pm 0.25\%$ in
beam momenta. On both wings of the $\Delta$, cross sections
vary rapidly with momentum; this introduces an error
several times larger than the errors quoted for cross sections.
We have added in quadrature to the experimental
errors an error allowing for this uncertainty in beam momentum.
This correction has been in the SAID data-base since 1980.
For reference, values of these errors are given in Table 1.

%Table 1
\begin{table}[htb]
\begin {center}
\begin{tabular}{cccc}
\hline
Lab Energy (MeV) &
$\sigma ^+_{tot}$ (mb) & Lab Energy (MeV) & $\sigma ^-_{tot}$
(mb)\\\hline
  71.6 & $26.09  \pm 0.62$ & 76.7 & $15.80 \pm 0.30$ \\
  97.4 & $54.68  \pm 0.66$ & 96.0 & $23.12 \pm 0.24$ \\
   -   &        -          &114.4 & $33.74 \pm 0.34$ \\
  118.9 & $96.25  \pm 1.25$ & 119.9 & $38.44 \pm 0.47$ \\
  120.4 & $101.04 \pm 1.53$ & 127.2 & $44.48 \pm 0.46$ \\
  136.0 & $141.19 \pm 1.42$ & 140.9 & $55.35 \pm 0.43$ \\
  138.7 & $148.60 \pm 1.42$ &  - & -  \\
  155.8 & $189.66 \pm 1.08$ & 159.6 & $67.90 \pm 0.32$  \\
  161.2 & $198.38 \pm 1.09$ & 164.7 & $70.26 \pm 0.33$ \\
  168.0 & $204.24 \pm 0.85$ & 172.6 & $70.74 \pm 0.33$ \\
  182.5 & $202.91 \pm 1.03$ & 184.6 & $69.76 \pm 0.35$ \\
  205.3 & $173.46 \pm 1.27$ & 208.9 & $59.43 \pm 0.43$ \\
  228.5 & $134.81 \pm 1.32$ & 232.6 & $47.06 \pm 0.42$ \\
  254.1 & $100.63 \pm 0.97$ & 255.4 & $38.46 \pm 0.29$ \\
  282.8 & $73.82 \pm 0.91$  & 290.1 & $29.97 \pm  0.26$ \\\hline
\end{tabular}
\caption{Total cross sections with errors including uncertainties in
beam momentum.}
\end {center}
\end{table}

\section {The essential discrepancy in data}
The present status of the phase shift analysis may be summarised
very simply.
Most data sets contribute close to $\chi ^2 = 1$ per point (when
analysed without the dispersion relation constraint).
There are few problems in fitting the {\it shapes} of differential
cross sections or polarisations.

However, there is a well known discrepancy in data in the mass range
below 145 MeV. One one side are the total cross sections $\sigma
^+_{tot}$ and $\sigma ^-_{tot}$ of Carter et al. [3] and the
normalisations of differential cross sections of Bussey et al. [4]. On
the other side are the normalisations of differential cross sections of
Brack et al. [12]. The shapes of differential cross sections from both
Bussey et al. and Brack et al. may both be fitted adequately, but there
are differences in normalisation.

There are faults here on both sides.
The $\pi ^-p$ total cross section measurement of Carter et al.
at the lowest energy 76.7 MeV is 7.5 standard deviations too high to
fit the shape of the $\Delta$.
That can be seen in Fig. 2(b) of the 1973 phase shift analysis of Carter
et al. [6]; the 76.7 MeV point lay well above an
effective range formula for $P_{33}$.
The next point at 96.0 MeV lies suspiciously high by 3.8 standard
deviations.
We therefore reject both these $\pi ^-p$ total cross sections at 76.7
and 96 MeV.
There is no difficulty for $\pi ^+p$.

It is a matter of conjecture where the problem lies with the
two $\pi ^-p$ points. The most likely explanation arises from
$\pi \to \mu$ decays in the region of the target.
At these low momenta, pions decay faster than they interact.
There is a Jacobean peak in the decay angular distribution at a
transverse momentum of 40 MeV/c.
Energy loss in the full target is greater than in the empty target,
increasing the decay rate.
A correction is needed for this in the extrapolation to zero solid
angle in the total cross section determination and may have been
underestimated.
The problem is worse for $\pi ^-$ than for $\pi ^+$ because
(a) elastic cross sections for $\pi ^-$ are lower than for $\pi ^+$
by about a factor 9,
(b) the beam size was somewhat larger for $\pi ^- $ than for $\pi ^+$.

Apart from this, there is one further discrepancy within
CERN Data.
At 263.7 MeV, the normalisation of Bussey et al. $d\sigma ^-/d\Omega$
data is $\sim 5\%$ too low to agree with the difference
between $\sigma ^-_{tot}$ and $\sigma ^0$. There is no
problem with the total cross section data at this energy
or with $\pi ^+p$ data.
It is therefore desirable to free the normalisation of
$d\sigma ^-/d\Omega$ at this single energy.

Otherwise, normalisations of Carter et al. total cross sections are
internally consistent with the integrated differential cross sections
of Bussey et al. added to $\sigma ^0$.
This provides a valuable check on resonance. There, partial waves other
than $P_{33}$ contribute $<5\%$ of the $\pi ^+p$ total cross section
and only $10\%$ of the $\pi ^-p$ total cross section. The small partial
waves are accurately determined from polarisation data via interference
with $P_{33}$. On resonance, the $P_{33}$ cross section is given by
$8\pi /k^2$, leaving no freedom in the absolute normalisation. The data
satisfy this check within experimental errors of typically
0.5\%.
Of course, it is still possible that normalisation
errors develop at lower momenta.
Incidentally, the data of Pedroni et al. [23] show
3 standard devations disagreements on resonance with this check,
despite their larger errors.

The relative normalisations of Brack et al.
$d\sigma ^\pm /d\Omega$ data are lower than those of
CERN data by amounts up to $\sim 10\%$ in the mass range below
140 MeV.

This normalisation discrepancy affects
primarily $P_{33}$ and hence $g^2_c$;
there are also small effects on the $\pi ^\pm p$ scattering lengths.
In order to estimate systematic errors arising from the choice of
data set, we consider two extremes.
In Fit I, the total cross sections  of Carter al al. are removed
and the normalisations of Bussey et al are floated;
data of Brack et al. are fitted according to published normalisations.
In fit II, the procedure is reversed: normalisations of Brack et al.
data and total cross sections of Pedroni et al. are floated;
the published normalisations of Bussey et al. are retained and the
total cross sections of Carter et al are fitted.
In both cases, charge exchange cross sections of Bugg et al. are
fitted, since these are the main source of information on $P_{13}$.
In both cases, normalisations of differential cross sections of
Frank et al.[24] and Bertin et al [25] are floated freely.
The normalisation of the Frank et al. differential cross sections
is low on average by 9\% and the normalisation of Bertin et al.
data is high on average by 12.4\%.

\section {Results}
In Fit I (without Carter et al. and Bussey et al. normalisations),
$g^2_c/4\pi$ optimises at 13.65. In fit II (floating Brack et al.
data), $g^2_c/4\pi$ optimises at 13.80. In both cases, purely
statistical errors are extremely small.
If one uses the difference between the two results as a guide to
systematic errors and attributes equal errors to both experiments,
the systematic error on each is $\sim \pm 0.10$.
Table 2 displays mean $\chi ^2$ per point for various data sets.

%Table 2
\begin{table}[htb]
\begin {center}
\begin{tabular}{ccc}
\hline
Data set & $\chi ^2$ fit I & $\chi ^2$ fit II \\\hline
  Bugg et al. $\sigma ^0$              & 0.52 & 0.87\\
  Bussey et al, $d\sigma ^+/d \Omega$  & 1.27 & 1.13\\
  Bussey et al, $d\sigma ^-/d \Omega$  & 1.49 & 1.93\\
  normalisation, Brack et al. $\pi ^+$   & 4.17 &  - \\
  normalisation, Brack et al. $\pi ^-$   & 2.74 &  - \\
  Brack et al. $d\sigma ^+/d \Omega$    & 1.76 & 1.02\\
  Brack et al. $d\sigma ^-/d \Omega$    & 2.50 & 1.91\\
  Pedroni et al. $\sigma ^+_{tot}$         & 1.51 &  - \\
  Pedroni et al. $\sigma ^-_{tot}$         & 2.30 &  - \\
  Carter et al, $\sigma ^+_{tot}$          &  -   & 1.38  \\
  Carter et al., $\sigma ^-_{tot}$        &  -   & 1.16  \\\hline
  Joram et al., $d\sigma ^+/d\Omega $     & 3.39 & 3.95  \\
  Joram et al., $d\sigma ^-/d\Omega $     & 2.08 & 2.26  \\
  Wiedner et al., $d\sigma ^-/d\Omega $     & 2.65 & 2.60  \\
  Hauser et al., $d\sigma ^0/d\Omega $     & 4.36 & 4.45  \\
  Gordeev et al., $d\sigma ^+/d\Omega $     & 5.71 & 5.74  \\
  Gordeev et al., $d\sigma ^-/d\Omega $     & 4.38 & 4.27  \\\hline

\end{tabular}
\caption{Mean values of $\chi ^2$ per point in fits I and II.}
\end {center}
\end{table}

The normalisations of Brack et al.
for $d\sigma ^+/d\Omega$ in fit I give a large $\chi ^2$
of 4.17 per energy, even though the constraints on normalisation from
the CERN data have been dropped. The problem is worst at 66.8 MeV
where the fitted normalisation is 7\% low, with a quoted error
of 2.0\%; at 86.8 MeV, it is low by 5\% with a quoted error
of 1.4\%.
What constrains these normalisations are the PSI X-ray data
at threshold.
In fit II, these discrepancies in normalisation
increase to 11\% and 7\% respectively.

This discrepancy is apparent from the phase shift analysis of
Fettes and Matsinos [26].
They analyse only data below 100 MeV and find a $\pi ^+p$ scattering
length $a_3 = (0.077 \pm 0.003)m_\pi ^{-1}$, compared with the value
$0.0885^{+0.0010}_{-0.0021}m_\pi ^{-1}$ from X-ray data.

The $\chi ^2 $ for the normalisation of Brack et al. $d\sigma
^-/d\Omega$ data is also quite high: 2.74 per energy.
The problems in
fit I are worst at 117.1 MeV, where the normalisation contributes a
$\chi ^2 $ of 15.85; at 66.8 and 87.5 MeV, it contributes 5.4 and 5.2
to $\chi ^2$. At 45 MeV, $\chi ^2$ for the $\pi ^-p$ differential cross
section is 8.30 per point. In Fit  II, where CERN data are used with
their published normalisations, the fitted normalisations of Brack et
al. $\pi ^-p$ data are in the range 0.86 to 0.95 for all energies from
117.1 MeV downwards.

It has often been remarked that $\sigma ^-_{tot}$ data of
Pedroni et al. lie systematically lower than those of
Carter et al. below the $\Delta$ resonance. On the other hand,
their errors are large and overlap almost everywhere with the
more precise data of Carter et al.

An escape route for the Brack et al. data is a possible
violation of charge independence, allowing
greater freedom. However, Meissner [27] estimates from
Chiral Perturbation Theory that violations of charge
independence in elastic scattering are unlikely to be above
1\%.
Meissner concludes that the 7\% violation of charge independence
proposed by Matsinos [28] appears unlikely.

The effect of this normalisation question is to make the
$\Delta$ slightly narrower when the data of Brack et al. are used.
That leads to a lower value of $g^2_c/4\pi$.
Fit I, favouring the Brack et al. data leads to
$g^2_c/4\pi = 13.65$; fit II, favouring CERN data, leads
to $g^2_c/4\pi = 13.80$.
However, this difference is now small because of the weight of
other data.
The latest analysis reported by Pavan at
Menu 2001 [21] uses both CERN and Brack data and
reports $g^2_c/4\pi = 13.69 \pm 0.07$;
this is slightly closer to fit I because of the larger number of
points in the Brack et al. data.

Apart from these normalisation questions, the lower half of Table 2
shows a number of data sets with high $\chi ^2$.
The data of Joram et al. [29] and Weidner et al. [30] are
mostly in the Coulomb interference region. The problem with the Joram
et al. data lies in the shape of the differential cross sections; for
Wiedner et al. data the problem is that the normalisation is 10\% high.

There are also high $\chi ^2$ for the data of Hauser et al. [31]
and Goreev et al. [32]. These are not normalisation problems:
it appears that errors have been underestimated. However, removing
these data from the fit has little effect.

\subsection {Errors from Coulomb barrier Corrections}
There is some model dependence in Coulomb barrier corrections.
Tromborg et al. include in their description of driving forces
only the dominant nucleon exchange.
Bugg [8] includes in addition $\sigma$, $\rho$ and $\Delta$
exchanges. His results are $\sim 25\%$ lower in magnitude.
Gashi et al. [33] evaluate corrections only up to 100 MeV.
All work is subject to some systematic uncertainty in the
$\sigma$ interaction.

As an estimate of systematic errors from Coulomb barrier
corrections, we take the difference between those of Tromborg et al.
and Bugg.
The former leads to the values of $g^2_c/4\pi$ quoted above; the latter
gives values of $g^2_c/4\pi $ which are systematically lower by 0.06.
Adding this error in quadrature to the systematic error arising from
choice of data set, the overall systematic error is about $\pm 0.12$
for $g^2_c/4\pi$.

\subsection {Scattering Lengths}
$S$-wave scattering lengths for fits I and II are shown in
Table 3. Errors are systematic and are about $0.003m^{-1}_\pi$.
Fit II is closer to the X-ray result: $2a_1 + a_3 =
0.2649 \pm 0.0024 m_\pi ^{-1}$.
The TRIUMF data try to pull the $\pi ^- p$ scattering length
to lower values. Both fits reproduce within errors the current algebra
result that the symmetric combination of scattering lengths is
zero.

%Table 3
\begin{table}[htb]
\begin {center}
\begin{tabular}{ccc}
\hline
 & Fit I & Fit II \\\hline
$a_3$ & -0.0847 & -0.0872 \\
$a_1$ & 0.170  & 0.173\\
$2a_1 + a_3$ & 0.255 & 0.259 \\
$a_1 - a_3$ &  0.255 & 0.260 \\
$a_1 + 2a_3$ & 0.001 & -0.001 \\\hline
\end{tabular}
\caption{Scattering lengths in units of $m^{-1}_\pi$.}
\end {center}
\end{table}

\subsection {Mass and Width Differences for the $\Delta$}
Fitted masses and widths of $\Delta ^0$ and $\Delta ^{++}$
and their differences are shown in Table 4 for both fits.
All these values are after applying the correction for
the Coulomb barrier.
The masses are evaluated where the $\pi ^\pm p$ phase
shifts go through 90$^\circ$.
The mass difference is consistent with twice the mass
difference between neutron and proton.
We remark also that Pedroni et al [23] use deuterium data to
find $\pi ^-n$ cross sections; they give
$M(\Delta ^-) - M(\Delta ^{++}) = 3.9$ MeV (no error quoted).

These differences  in mass and width are visible by eye in the total
cross sections of Carter et al.; an illustration of the difference in
the Chew Low plot between $\pi ^+ p$ and $\pi ^-p$ is shown in Fig.
2(b) of Ref. [6]. However, the differences in mass and width
are also required in fit I and are therefore clearly required also by
differential cross section and polarisation data.

% Table 4
\begin{table}[htb]
\begin {center}
\begin{tabular}{ccc}
\hline

 & Fit I & Fit II \\\hline
 $M(\Delta ^{++})$ & $1231.45 \pm 0.3$ MeV & $1231.0 \pm 0.3$ MeV \\
 $M(\Delta ^0)$ & $1233.6 \pm 0.3$ MeV & $1232.85 \pm 0.3$ MeV \\
$M(\Delta ^0) - M(\Delta ^{++}) $ & $1.86 \pm 0.4$ MeV &$2.16 \pm 0.4$
MeV \\
$\Gamma (\Delta ^{++})$ & $114.8 \pm 0.9$ MeV & $115.0 \pm 0.9$ MeV \\
$\Gamma (\Delta ^0)$ & $116.4 \pm 0.9$ MeV & $118.3 \pm 0.9$ MeV \\
$\Gamma (\Delta ^0) - \Gamma (\Delta ^{++})$ & $1.6 \pm 1.3$ & $3.3
\pm 1.3$ \\\hline
\end{tabular} \caption{Masses and widths for the $\Delta$ from  fits I
and II.}
\end {center} \end{table}

Differences in width evaluated where the $\pi ^\pm p$ phase shift
goes through 90$^\circ$ are $1.16 \pm 1.3$ MeV for fit I
and $3.3 \pm 1.3$ MeV for fit II.
We now examine how to account for these differences in width.
To first approximation, the width may be parametrised as
proportional to $k^3/(1 + k^2R^2)$, where $k$ is the decay momentim
and $R$ is an effective radius for the centrifugal barrier,
taken to be 0.8 fm.
Because $k$ is larger in $\Delta ^0 \to n\pi ^0$ than for decay
to $\pi ^-p$, the width is larger. Allowing for the branching ratio
$2:1$ to these channels, one expects this phase
space difference to make $\Gamma ^0$ larger than $\Gamma ^{++}$
by 0.8 MeV for a given $\pi p$ mass. However, in Table 4,
$\Gamma ^{0}$ is evaluated at a mass which is higher by 2 MeV
than that for $\Gamma ^{++}$. This contributes a further 2.6 MeV to
$\Gamma ^0$. Thirdly, the $\Delta ^0$ has an extra width of 1.1
MeV for the $\gamma n$ channel. Adding these three effects, the
expected width difference is 4.3 MeV. This estimate is close to
Pilkuhn's estimate of 4.6 MeV [34].
These estimates are consistent with the
observed difference in widths in fit II.
We remark that the $P_{33}$ phase shift is not sensitive to
the total cross section near resonance. It is mostly sensitive
near half-height of the resonance, when the $P_{33}$ phase shift
is close to 45 or $135^\circ$. Hence the width is sensitive to
the formula used to parametrise the phase shift as a function
of mass. Here, simple spline fits are used in the SAID program.

Kruglov [35] reports an independent
Gatchina analysis of $\pi N$ partial waves with the results:
\begin {eqnarray}
M^0 &=& 1233.1 \pm 0.3~{\rm MeV} \\
M^{++} & = & 1230.5 \pm 0.2~{\rm MeV} \\
M^0 - M^{++} & = & 2.6 \pm 0.4~{\rm MeV} \\
\Gamma ^0 - \Gamma ^{++} & = & 5.1 \pm 1.0~{\rm MeV}.
\end {eqnarray}
From their original total cross section data, Carter et al.
found $\Gamma ^0 - \Gamma ^{++} = 6.4 \pm 1.8$ MeV.
We consider all these values consistent in view of the
different formulae used in different analyses.

\section {Discrepancies with the Goldberger-Treiman relation}
We return now to the Goldberger-Treiman relation, eqn. (1).
Our general approach follows the ideas of chiral symmetry
and its spontaneous breaking [36,37]. Our calculation
of the discrepancy with the Goldberger-Treiman relation
follows an algebraic approach proposed by Coon and Scadron [38].

The $\pi N$ pole in the $B^{(+)}$ dispersion relation is at
$s = m^2_n$, corresponding to $q^2 = 0$ for the mass of the pion.
However, the pion decay constant $f_\pi$ is determined at
$q^2 = m^2_\pi$.
It is necessary to allow for possible $q^2$ dependence of $f_\pi$.
The value of $g_A$ is determined at $q^2 \simeq 0$ because of the
low energy available in neutron beta decay.

The $q^2$ dependence of $f_\pi$ may be obtained from a
once-subtracted dispersion relation:
\begin {equation}
f_\pi (q^2) - f_\pi (0) = \frac {q^2}{\pi } \int ^\infty _0
\frac {dq'^2}{q'^2} \frac {Im f_\pi (q'^2)}{q'^2 - q^2}.
\end {equation}
We suppose that the pion decays to two constituent quarks which in
turn couple to $W^\pm$.
For the quark loop,
\begin {equation}
Im f_\pi (q^2) = \frac {3g_{\pi q\bar q}}{2} \frac {4\hat m}{8\pi }
\left( 1 - \frac {4\hat m^2}{q^2}\right)^{1/2} F_\pi (q^2 - 4\hat m^2)
~\theta (q^2 - 4\hat m^2).
\end {equation}
This follows from unitarity
with the inclusion of a factor 3 for colour;
$\hat m = (m_u + m_d)/2$ and $m$ are constituent quark masses.
At the quark level, the Golderger-Treiman relation is
$f_\pi g_{\pi q \bar q} = \hat m$.
Then, using a Taylor series expansion of the term
$1 - q^2/q'^2$ in the denominator, eqn. (6) becomes
$$ \frac {f_\pi (q^2) - f_\pi (0)}{f_\pi (0)} = \frac {m_\pi^2}{\pi }
\int ^\infty _{4\hat m^2} \frac {dq'^2}{q'^4}
\frac {3g^2_{\pi q\bar q}}{4\pi}
\left( 1 - \frac {4\hat m^2 }{q'^2} \right)^{1/2}
\left( 1 + \frac {q^2 }{q'^2} + \ldots \right)
F_\pi (q'^2 - 4\hat m^2).$$
With $F_\pi = 1$, this may be evaluated analytically:
\begin {equation} \Delta = \frac {f_\pi (m_\pi ^2) - f_\pi (0)}{f_\pi
(0)} = \frac {m^2_\pi }{8\pi ^2 f^2_\pi } \left(1 + \frac {m^2_\pi}{10
\hat m^2}\right). \end {equation} The first term of the result is
independent of $\hat {m}$. This term alone predicts a discrepancy
$\Delta = 0.0295$; using $g_c = m_ng_A/[f\pi (1 - \Delta)]$, we find
$g^2_c/4\pi = 13.99$. The second term in (8) is very small and it is
adequate to take $\hat m = m_n/3$. It leads to $\Delta = 0.0301$ and a
prediction \begin {equation} g^2_c/4\pi  = 14.01 \pm 0.10,
\end{equation}
where the error arises from uncertainties in $f_\pi$ and $g_A$.

For the form factor $F_\pi$, we use
$F_\pi  = \exp (-k^2r_\pi^2/6)$, where $k$ is the momentum
in the $q\bar q$ loop.
The pion charge radius is given [39] by $r_\pi^2 = 3/(4\pi ^2 f^2_\pi )
= (0.59~fm)^2$, in close agreement with the experimental value of 0.67
fm.
Vector Dominance predicts $r_\pi = \sqrt {6}/m^2_\rho = 0.63~fm$. We
use this intermediate value. Folding in $F_\pi (q^2)$ numerically gives
a factor 0.674 multiplying the right-hand side of eqn. (8). Finally,
the prediction  is \begin {equation} g^2_c /4\pi = 13.74 \pm 0.10; \end
{equation} this lies in the middle of the experimental values.

The value of $g^2_0/4\pi$ may also be evaluated by using in
eqn. (8) the mass of the $\pi ^0$. The result is $g^2_0/4\pi =
13.70 \pm 0.10$;
this is 1.5 standard deviations above the value of Stoks et al.,
$13.56 \pm 0.09$.

A number of other determinations of the discrepancy $\Delta$
have been made by less direct methods [40-43] but with similar
results.

\section {Conclusions}
>From the $B^{(+)}$ unsubtracted dispersion relation,
accurate values of $g^2_c/4\pi$ may be determined.
The limitation at present is the systematic discrepancy
between CERN and TRIUMF data below 145 MeV,
although the addition of other data today reduces the
effect of the discrepancy to quite a small value.
The CERN data prefer $g^2_c/4\pi = 13.80 \pm 0.12$
and the TRIUMF data $g^2_c/4\pi = 13.65 \pm 0.12$.
These errors include systematic uncertainties of $\pm 0.06$
from uncertainties in the Coulomb barrier corrections.
Normalisations of TRIUMF data are poorly fitted even when
the normalisations of CERN data are floated.

The prediction from the Goldberger-Treiman relation,
after allowing for the $q^2$ dependence of $f_\pi$ from
chiral symmetry, is $g^2_c/4\pi = 13.74 \pm 0.10$.

The mass difference between $\Delta ^0$ and $\Delta ^{++}$ is
consistent with twice the mass difference between neutron and
proton.
The observed difference in width is consistent with the effects of
phase space and a 1.1 MeV width difference due to the extra channel
$\Delta ^0 \to \gamma n$.

\section {Acknowledgements}
We wish to thank Richard Arndt and Marcello Pavan for
assistance in modifying the SAID program to our needs
and for help in getting it running.

\end {document}